\pdfoutput=1

\documentclass[rsc,pre,preprint,superscriptaddress]{revtex4-1}

\usepackage[caption=false]{subfig}
\usepackage{graphicx}
\usepackage{amsmath}
\usepackage{amssymb}
\usepackage[hidelinks]{hyperref}
\usepackage[normalem]{ulem}
\usepackage{siunitx}
\usepackage{float}
\usepackage{cleveref}
\floatstyle{plaintop}
\restylefloat{table}

\newcommand*\diff{\mathop{}\!\mathrm{d}}

\usepackage{color}

\begin{document}

\title{Transient response and domain formation in electrically deforming liquid crystal networks}

\author{Guido L. A. Kusters}
\email[]{g.l.a.kusters@tue.nl}
\affiliation{Department of Applied Physics, Eindhoven University of Technology, The Netherlands}


\author{Paul van der Schoot}
\affiliation{Department of Applied Physics, Eindhoven University of Technology, The Netherlands}

\author{Cornelis Storm}
\affiliation{Department of Applied Physics, Eindhoven University of Technology, The Netherlands}
\affiliation{Institute for Complex Molecular Systems, Eindhoven University of Technology, The Netherlands}

\date{\today}

\begin{abstract}
Recently, Van der Kooij and co-workers recognized three distinct, transient regimes in the dynamics of electrically-deforming liquid crystal networks [Van der Kooij et al., Nat. Commun. \textbf{10}, 1 (2019)]. Based on a Landau-theoretical framework, which encompasses spatially resolved information, we interpret these regimes: initially, the response is dominated by thermal noise, then the top of the film expands, followed by a permeation of this response into the bulk. An important signature of this interpretation is a significant dependence of the regime time scales on film thickness, where we observe a clear thin-film-to-bulk transition. The point of transition coincides with the emergence of spatial inhomogeneities in the bulk, i.e., domain formation, and should be avoided due to the less predictable steady-state expansion it gives rise to. Finally, we show that this domain formation can be suppressed by decreasing the initial thickness of the film, and increasing the linear dimensions of the mesogens, or their orientational order when crosslinked into the network, though this comes at the cost of the deformation magnitude. Our results contribute to achieving finer control over how smart liquid crystal network coatings are activated.
\end{abstract}

\pacs{aaa}

\maketitle

\section{Introduction}\label{sec:introduction}
Liquid crystal networks are composite materials comprising liquid-crystalline mesogens embedded in a polymer matrix, similar to liquid crystal elastomers \cite{finkelmann1981investigations}. In both materials, a phase transition of the liquid crystal can be effected by applying an external stimulus, such as a temperature variation \cite{cao2019temperature,babakhanova2019surface}, UV irradiation \cite{white2012light,stumpel2014stimuli}, or an electric field \cite{liu2017protruding,van2019morphing,van2020electroplasticization}, which grants mechanical control over their deformation through a coupling to the polymer network. In the case of liquid crystal elastomers, this coupling has already been exploited in the form of, e.g., soft actuators \cite{wermter2001liquid,ikeda2007photomechanics,sanchez2009photo,sanchez2011liquid,dai2013humidity,schuhladen2014iris,zeng2017self}, shape-memory materials \cite{rousseau2003shape,liu2007review,burke2010soft,burke2013evolution,kotikian20183d}, and haptic feedback \cite{campo2011nano,camargo2011microstamped,camargo2011localised,camargo2012batch,torras2014tactile}.

Although the same basic coupling underlies liquid crystal networks, and they consist of the same molecular building blocks as liquid crystal elastomers, albeit with a significantly larger density of both permanent crosslinks and mesogens, these materials are more suitable to slightly different ends. This is because their response turns out to be markedly different, strikingly being able to generate and sustain changes of volume (of up to ten percent) upon actuation \cite{liu2017protruding,van2019morphing,van2020electroplasticization}, whereas conventional liquid crystal elastomers are known to deform at approximately constant volume. This suggests that liquid crystal networks allow for potentially even richer applications in the field of smart coatings, e.g., with controllable surface topographies \cite{liu2012photo,mcconney2013topography,liu2017protruding,babakhanova2019surface,van2019morphing,van2020electroplasticization}, adaptive adhesion and friction \cite{liu2014self,gelebart2018photoresponsive}, and control over the transport of molecular cargo \cite{cao2019temperature,zhan2020localized}.

\begin{figure*}[htbp]
	\centerline{
		\includegraphics[width=16cm]{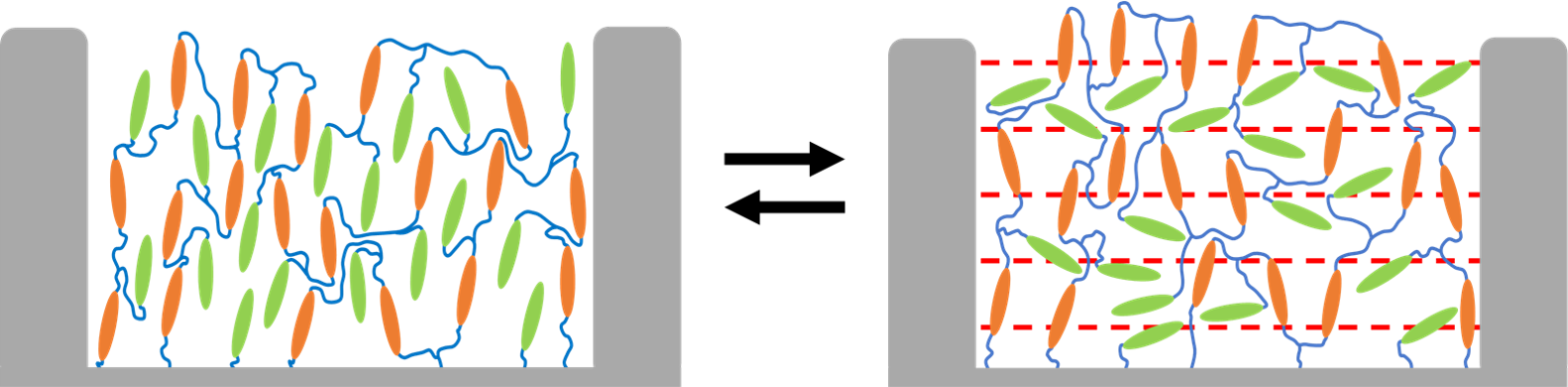}
	}
	\caption{Schematic representation of the liquid crystal network film, sandwiched between two gray electrodes, in the absence (left) and presence (right) of an electric field (dashed red lines). Cross-linked mesogens are indicated in orange, pendant (dipolar) mesogens are indicated in green and polymer strands are indicated in blue. The mesogens are prepared with homeotropic alignment and the black arrows indicate the transition between the ``off" state (left) and the ``on" state (right) of the liquid crystal network, which can be tuned reversibly. Figure adapted from Ref. \cite{kusters2021permeation}.}
	\label{fig:diagram geometry}
\end{figure*}

For such applications to be practically implemented, one key question that must be addressed regards the time scales on which they can be activated, i.e., the transient response that must occur before a steady state is reached. This question is especially relevant in light of the recent experimental work of Van der Kooij and co-workers \cite{van2019morphing}, who consider an electrically-deforming liquid crystal network comprising two distinct species of mesogen (see Figure \ref{fig:diagram geometry}): (i) mesogens that form permanent crosslinks in the polymer network (orange), and (ii) end-on grafted side-group mesogens, functionalized with a strong, permanent dipole moment (green); due to the crosslinking into the polymer network, the former are much less mobile than the latter. In their experiments, the mesogens are prepared on a substrate with homeotropic alignment, and an alternating electric field (red) is subsequently applied in the perpendicular direction, i.e., in the plane. Here, typical length scales for the liquid crystal network film are $\sim\SI{2.5}{\micro\meter}$ for the thickness and $\sim0.5\times\SI{0.5}{\milli\meter}$ for the activated area; the applied electric field strengths lie in the range $\sim1-\SI{10}{\volt\per\micro\meter}$.

Van der Kooij and co-workers report that, upon actuation, the liquid crystal network invariably undergoes three distinct, temporally separated regimes before the steady state, in which significant volume increases of up to ten percent can be sustained, is reached. Strikingly, the time scales on which these regimes occur are significantly slower than that of the actuation, with typical oscillation frequencies being in the range of hundreds of kilohertz, and the material generally only entering the steady state on the order of tens of seconds. They provide a phenomenological hypothesis by associating the different regimes with the sequential response of the side-group and the crosslinked mesogens, followed by the progressive weakening of the polymer matrix due to the coherent oscillations of the mesogens. Since the existence of these regimes directly limits the response time of electrically deforming liquid crystal networks, in this paper we further investigate their transient response theoretically.




In prior work, Warner, Terentjev, and co-workers pioneered a theoretical framework combining the classical theory of elasticity with the Landau-de-Gennes theory of liquid crystals, in which the liquid crystal elastomer effectively explores a random walk, rendered anisotropic by the presence of liquid-crystalline mesogens \cite{warner1988theory,warner1991elasticity}. This well-celebrated, ``neo-classical" framework explains, among other things, the occurrence of phase transitions, instabilities, and (quasi)soft modes of deformation \cite{bladon1993transitions,verwey1996elastic,tajbakhsh2001spontaneous,golubovic1989nonlinear,warner1994soft,olmsted1994rotational,verwey1995soft,martinoty2004mechanical,terentjev2004commentary,stenull2004commentary,martinoty2004replyT,martinoty2004replyS,menzel2009response}.


For our purposes in this paper, however, we require a different approach. This is because the polymer strands connecting the mesogens in a liquid crystal network are too short and too densely constrained to effectively explore a random walk, meaning the foundational principles of the neo-classical theory no longer apply. Furthermore, even if we were to nevertheless apply this framework, and adapt it slightly to allow for non-volume-preserving deformations, the predicted changes in volume are much too small to explain the experimental observations \cite{Thesis}.

To address this, we propose a dynamical Landau-de Gennes theory, based primarily on the state of the mesogens, rather than on the configuration of the polymer network; the motivation for this choice is that the liquid crystal network is extremely dense in mesogenic component, so as to effectively be liquid crystalline itself. In a series of recent papers, we already obtained promising qualitative agreement between the theory, molecular dynamics computer simulations, and experiments, and have also characterized how in a thin-film geometry, as is common in experiments, the response to a constant electric field permeates the liquid crystal network from top to bottom \cite{kusters2020dynamical,kusters2021permeation}. 

In the current paper we shall use this theoretical framework to investigate the response of thin-film liquid crystal networks to alternating electric fields, focusing specifically on the transient response; this allows us to probe the time scales on which the material can be activated in practice. To this end, in Sec. \ref{sec:model}, we provide a more detailed description of our model system, inspired by the experiments of Van der Kooij and co-workers \cite{van2019morphing}, and we introduce the relevant theory, in Secs. \ref{sec:theory}-\ref{sec:scaling}. 

Then, in Sec. \ref{sec:results}, we illustrate the typical response of the liquid crystal network to an alternating electric field, focusing on the transient dynamical behavior. In particular, we identify three characteristic time scales, associated with an initial response suppressed by thermal noise, followed by expansion of the top of the film, and finally the bulk of the material. This provides an alternative explanation for the reported experimental findings, and suggests spatial information translates directly to distinct dynamical behavior on an experimentally observable scale. 

Following this, in Sec. \ref{sec:results2}, we study how these time scales vary with the thickness of the film, and recover a clear thin-film-to-bulk transition, characterized by the onset of domain formation. We argue that prospective applications should be operated either in the thin-film limit, where relatively few domains are formed, or in the bulk limit, where domain walls are small relative to the size of the material; near the transition point, both effectiveness and reliability suffer. Furthermore, in Sec. \ref{sec:results3}, we contextualize our results by quantifying the number of formed domains as a function of model parameters. We find that domain formation can be suppressed, and thus steady-state actuation accelerated, by decreasing the initial thickness of the film, increasing the linear dimensions of the mesogens, and increasing the degree of orientational order with which the mesogens are crosslinked into the network. This comes at the cost of a decreased deformation magnitude, however. Finally, we summarize our most salient findings and critically compare our results to the experiments of Van der Kooij and co-workers \cite{van2019morphing}, in Sec. \ref{sec:conclusion and discussion}.

\section{Model}\label{sec:model}
Inspired by the experimental work of Refs. \cite{liu2017protruding,van2019morphing,van2020electroplasticization}, we consider as our model system a liquid crystal network of densely connected mesogens. Experimentally, such a network is realized by synthesizing di-acrylates of the forms (i) =oligomer-mesogen-oligomer= and (ii) =oligomer-mesogen in approximately equal number; no additional polymer strands are introduced in this procedure. This ensures that virtually all mesogenic units are closely interconnected within the network by means of the short polymer strands incorporated in the di-acrylate ingredients. This also implies that the resulting network does not typically conform to either the classical main- or side-chain geometries (see Fig. \ref{fig:diagram geometry}), and that macroscopic phase separation of the polymer and mesogenic components is prohibited. Given that the extent of the used polymer strands is on the scale of the mesogen length, the liquid crystal network is more liquid crystal-like than network-like, with the distribution of mesogens largely determining the mechanical properties of the material.

As the synthesis ingredients suggest, the mesogens in our model system come in two distinct species: (i) mesogens that are fully crosslinked into the network on either end, and (ii) pendant side-group mesogens that are only connected to the network on one end. The former (orange in Fig. \ref{fig:diagram geometry}) arguably are relatively immobile, as corroborated by molecular dynamics computer simulations \cite{kusters2020dynamical}, and thus unable to respond appreciably if an electric field is applied. Conversely, the latter (green in Fig. \ref{fig:diagram geometry}) are much more mobile and interact strongly with an applied electric field by virtue of being functionalized with a strong, permanent dipole moment. Since the liquid crystal network is effectively liquid crystalline itself, the ratio of crosslinked-to-dipolar mesogens used largely determines the crosslinking density of the material; this we do not vary for the purpose of this paper. In the experimental system, the mesogens are attached to a substrate with homeotropic alignment, and form a thin film. During actuation, an alternating electric field is then applied in the transverse direction, i.e., in the plane. As noted, this can result in sustained volume increases of up to ten percent.

To describe this system, motivated by the crucial role played by the mesogens, we propose a Landau-type theory that is based on the state of the mesogens, and how this couples to the volume of the liquid crystal network, rather than on (neo-)classical polymer elasticity. Although, due to the generic nature of our model, it encompasses various underlying mechanisms, ranging from an electrically-driven glass transition to the collective motion of an array of coupled oscillators \cite{kusters2021permeation}, for the purpose of this paper we shall illustrate it based on excluded-volume arguments. That is, upon application of the electric field, the pendant, dipolar mesogens reorient in response to it, whereas the crosslinked mesogens remain relatively immobile. This increases their mutually-excluded volume, and in turn the total volume of unoccupied space in the liquid crystal network; the collective effect of many such reorientations effects a macroscopic expansion of the material. 

In practice, steady-state operation of these liquid crystal networks requires alternating electric fields, as the polymeric material continually rearranges to fill the generated pockets of free volume. To emulate this balance in the model, we include, in addition to the above, a phenomenological, viscoelastic relaxation of the liquid crystal network volume. The resulting model coincides with the theory we previously reported in Ref. \cite{kusters2021permeation}.

In the following sections we briefly discuss the theory, which we shall use for the analysis presented in the remainder of this paper. Readers solely interested in our results can safely skip to Sec. \ref{sec:results}, where we discuss the resulting transient dynamics.

\section{Equilibrium theory}\label{sec:theory}
We describe the liquid crystal network in terms of two key order parameters: one characterizing the orientational order of the mesogens and the other specifying the concomitant volume expansion. The former of these is the driving force in the model. In particular, since only one species of mesogen responds appreciably to an applied electric field, the other being fully crosslinked into the network, we suffice by considering only the reorientations of pendant, dipolar mesogens against a background of immobile crosslinked mesogens. In this context, we carried out Molecular Dynamics (MD) computer simulations indicating that, upon application of an electric field, the dipolar mesogens generally either (i) reorient to align with the electric field, perpendicular to the crosslinked mesogens, or they (ii) are impeded in their orientation due to the (excluded-volume) interactions with the crosslinked mesogens, remaining along their initial axis or orientation \cite{kusters2020dynamical}. This motivates us to introduce a two-population model in terms of the order parameter $f$ [$-$], where $0\leq f^2\leq1$ is the fraction of electric-field-aligned dipolar mesogens. Note that we write $f^2$, rather than $f$, to ensure a consistent interpretation as a (positive) fraction.

The above suggests the model expresses a competition between the electric field, quantified by the field strength $H$ [\SI{}{\joule/\meter^3}] \footnote{The form $H\propto\lvert\underline{E}\rvert^2$ is appropriate for describing the free energy of an induced dipole due to the inversion symmetry of the nematic director $\underline{n}\rightarrow-\underline{n}$. Although for a mesogen with permanent dipole moment $\underline{p}$ the analogous expression would read $-\underline{p}\cdot\underline{E}$, averaging this over an ensemble of mesogens using a Boltzmann distribution this can again be written in the form $-\langle\underline{p}\cdot\underline{E}\rangle\propto\lvert\underline{p}\rvert^2\lvert\underline{E}\rvert^2$ for sufficiently weak electric fields, $\lvert\underline{p}\rvert\lvert\underline{E}\rvert\ll k_BT$ \cite{prost1995physics,de2012liquid,vertogen2012thermotropic}. Although this relationship becomes linear in the limit of strong electric fields, this does not alter our qualitative conclusions in this paper}, favoring reorientation of the dipolar mesogens along the electric-field axis on the one hand, and the (excluded-volume) interactions with the crosslinked mesogens, favoring the initial axis of orientation, on the other hand. The latter can be quantified by a critical field strength $H_*$ [\SI{}{\joule/\meter^3}], implying there exists a threshold value for the electric field strength above which reorientation of the dipolar mesogens becomes energetically favorable. In fact, such a critical field strength emerges naturally from a more elaborate derivation of the theory, as discussed in Ref. \cite{kusters2020dynamical}, and can be shown to depend on the linear dimensions of the mesogens, their orientational order at the time of crosslinking and the crosslinking fraction of the network. The Gibbs free energy per unit \textit{reference} volume can then be written as \footnote{If we were to write down the \textit{actual} Gibbs free energy density, $\mathcal{G}=G/V$, with $G$ the total Gibbs free energy, thermodynamic equilibrium, i.e., minimization of $G$ with respect to $\eta$, would demand $\partial \mathcal{G}/\partial\eta=-\mathcal{G}/\left(1+\eta\right)$. This means the Gibbs free energy density must take the form $\mathcal{G}=g/\left(1+\eta\right)$, with $\partial g/\partial \eta=0$. This indicates that, if we wish to treat $\eta$ as a proper order parameter, the Gibbs free energy per unit \textit{reference} volume, $g= G/V_0$, is the relevant thermodynamic potential. We explicitly \textit{construct} the Landau theory to yield the proper behavior upon minimization with respect to all order parameters}
\begin{equation}\label{eq:g1}
    g_1=\frac{1}{2}\left(H_*-H\right) f^2+\frac{1}{4}B_ff^4,
\end{equation}
where $B_f$ [\SI{}{\joule/\meter^3}] denotes a bulk-modulus-like coefficient that tempers mesogen reorientation, as this induces local strains in the polymer matrix. Although, in this case, such a term is not strictly required to bound the free energy from below, provided we manually demand that $0\leq f^2\leq1$. This ensures that a minimization of Eq. \eqref{eq:g1} can produce intermediate order parameter values $0<f^2<1$, representing alignment of a fraction of dipolar mesogens.

Following this, we introduce the volume-expansion order parameter $\eta=\left(V-V_0\right)/V_0$ [$-$], with $V$ and $V_0$ the current and initial system volume, respectively. Expansion of this volume is driven by the mutually-excluded volume of the mesogens, meaning that an increased fraction of dipolar mesogens aligned with the electric field, i.e., perpendicular to the axis of orientation of the crosslinked mesogens, favors an increase in system volume. Neglecting, for now, the viscoelastic relaxation of the network (we shall return to this below when we discuss our model dynamics), and taking $\xi$ [\SI{}{\joule/\meter^3}] as the coupling coefficient, we supplement Eq. \eqref{eq:g1} with
\begin{equation}\label{eq:g2}
    g_2=-\xi\eta f^2+\frac{1}{2}B_\eta \eta^2.
\end{equation}
The bulk-modulus-like term proportional to $B_\eta$ [\SI{}{\joule/\meter^3}] is now formally required to keep the free energy bounded from below. We refrain from adding an explicit pressure-volume contribution, as the role of pressure can effectively be absorbed in model parameters (not shown). 

Next, we integrate the total Gibbs free energy per unit \textit{reference} volume, $g=g_1+g_2$, over the initial volume of the liquid crystal network. In our case, this concerns a thin film extending from the reference coordinate $z_0=0$ to $L_0$, the initial thickness of the film, to yield the Gibbs free energy \cite{kusters2021permeation}
\begin{equation}\label{eq:Gibbsreference}
    \frac{G}{A}=\int_0^{L_0}\diff z_0 \left[g+\frac{\kappa_f^2}{2\left(1+\eta\left(z_0\right)\right)^2}\left(\frac{\partial f\left(z_0\right)}{\partial z_0}\right)^2
    +\frac{\kappa_\eta^2}{2\left(1+\eta\left(z_0\right)\right)^2}\left(\frac{\partial\eta\left(z_0\right)}{\partial z_0}\right)^2\right],
\end{equation}
with $A$ the lateral area of the liquid crystal network. Here, we have added square-gradient contributions proportional to the phenomenological coefficients $\kappa_f$ [\SI{}{\sqrt{\joule/\meter}}] and $\kappa_\eta$ [\SI{}{\sqrt{\joule/\meter}}], to promote smooth spatial profiles of the corresponding order parameters. Generally, such coefficients derive from local interactions \cite{de1979scaling}, and so are not fully independent from other model parameters. Finally, the prefactor $1/\left(1+\eta\left(z_0\right)\right)^2$ makes explicit the geometric significance of the volume-expansion order parameter, $\eta\left(z_0\right)$, in locally smearing out such spatial gradients.

The equilibrium state of the liquid crystal network can now be recovered by functionally minimizing Eq. \eqref{eq:Gibbsreference} with respect to the order parameters $f\left(z_0\right)$ and $\eta\left(z_0\right)$, subject to the following boundary conditions. At the bottom of the film, the mesogens are clamped to the substrate, and so remain along their initial axis of orientation regardless of the applied electric field. Since a lack of reorientation in turn precludes local volume expansion, we impose $f\left(0\right)=\eta\left(0\right)=0$.

At the top of the film, we model a diffuse interface of the liquid crystal network with the ambient medium. To this end, we assume that the density varies continuously over the interface, such that the volume-expansion order parameter $\eta\left(z_0\right)$, which is proportional to an inverse density, assumes its maximum value at $z_0=L_0$. We compute this saturated expansion by minimizing $g_2$, taking heed of the restriction $0\leq f^2\leq1$, which has us set $\eta\left(L_0\right)=\xi/B_\eta$. Although the increased expansion near the top of the film makes it easier for those mesogens to locally reorient, we need not impose this as a boundary condition for $f\left(z_0\right)$; this is already incorporated in the couplings of the model (see Eq. \eqref{eq:g2}). Instead, we demand the reflecting boundary condition $\partial_{z_0}f\left(L_0\right)=0$, which prohibits order parameter flow through the top of the film. This boundary condition is slightly different from the one we used in Ref. \cite{kusters2021permeation}; we shall return to this point in Sec. \ref{sec:conclusion and discussion}.

This concludes the discussion of our equilibrium theory. Below, in Sec. \ref{sec:dynamics}, we explain how we extend this framework to also study the dynamical behavior of the liquid crystal network, paving the way for an analysis of the transient dynamics.

\section{Relaxational dynamics}\label{sec:dynamics}

To describe the temporal evolution of our order parameters, which we treat as non-conserved \footnote{It is also possible to describe the volume-expansion order parameter, $\eta$, as a conserved order parameter. In that case, we would need to impose an influx of volume at the upper boundary of the film to ensure expansion remains possible. The reason we opt to treat $\eta$ as non-conserved is because we expect the time scales associated with the transport of free volume to be much shorter than the other time scales under consideration in this paper. If such transport is effectively instantaneous, the above reduces to a non-conserved treatment}, the simplest form of dynamical equations we can supplement Eq. \eqref{eq:Gibbsreference} with describes relaxational dynamics, according to
\begin{equation}\label{eq: dynamics}
\begin{split}
    \partial_{t}{f}&=-{\Gamma}_{f}\frac{\delta {G}}{\delta {f}}+\theta_f,\\
    \partial_{t}{\eta}&=-{\Gamma}_{{\eta}}\frac{\delta {G}}{\delta {\eta}}+\theta_\eta-{\gamma} \, \left({\eta}-\eta_0\right) \, \tau\left(z_0,t\right).
\end{split}
\end{equation}
Here, $\Gamma_f$ [\SI{}{\meter^3/\joule \second}] and $\Gamma_\eta$ [\SI{}{\meter^3/\joule \second}] denote kinetic coefficients that encompass the dissipative processes through which the free energy may be lowered by variation of the relative order parameter. The Gaussian noise terms $\theta_f$ [\SI{}{1/\second}] and $\theta_\eta$ [\SI{}{1/\second}] subsequently ensure that the system eventually approaches the global minimum of the free energy, and we choose their properties such that the fluctuation-dissipation theorem is satisfied \cite{glauber1963time}. In particular, the Gaussian noise terms enable the system to depart from its initial configuration, which is marginally stable. The set of dynamical equations is completed with the addition of a term reflecting the viscoelastic relaxation of the polymer network, which we neglected in the previous section. This term, proportional to the phenomenological coefficient $\gamma$ [\SI{}{1/\second^2}], guarantees that the volume-expansion order parameter profile, $\eta\left(z_0,t\right)$, invariably relaxes back to its initial profile, i.e., the equilibrium profile it assumes in the absence of an electric field, $\eta_0\left(z_0\right)$, as a function of time $t$. This presupposes an enforced equilibrium of the model, imposed on top of the free energy functional of Eq. \eqref{eq:Gibbsreference}, which is informed by experiments and Molecular Dynamics computer simulations \cite{liu2017protruding,van2019morphing,van2020electroplasticization,kusters2020dynamical}.

To achieve this, it is insufficient to simply write the relaxational term as $-\gamma\left(\eta-\eta_0\right)$. Although this indeed induces exponential relaxation of the volume-expansion order parameter profile, this relaxation occurs toward an equilibrium profile satisfying $\gamma\left(\eta-\eta_0\right)=\Gamma_\eta\frac{\delta {G}}{\delta {\eta}}$, rather than toward the initial profile $\eta_0\left(z_0\right)$. Although we can resolve this point by making the relaxational term an increasing function of the time $t$, such that the relaxational term dominates in the long-time limit, this introduces a different problem. Namely, if we apply an alternating electric field to the liquid crystal network, we expect this to eventually result in a steady-state expansion of the liquid crystal network, as reported in experiments \cite{liu2017protruding,van2019morphing,van2020electroplasticization}. However, a relaxational term that increases monotonically with $t$ suppresses \textit{any} volume expansion if we wait long enough, rendering a dynamic steady state impossible in the long term. Thus, we require a more sophisticated relaxation function, $\tau\left(z_0,t\right)$ [\SI{}{\second}], which locally keeps track of the deformation history of the liquid crystal network to determine how strongly a given volume element should relax \cite{kusters2021permeation}. This is a highly non-trivial component of the theory, and although it is not a true memory kernel, it fulfills a similar function. 


As the basis for this relaxation function we take $\tau\left(z_0,t\right)=t$, which describes Gaussian relaxation. This is the simplest functional form that invariably achieves relaxation of the volume-expansion order parameter to its initial profile in the long-time limit. In order to circumvent the suppression of volume expansion in this limit, we alter this function by following the philosophy that the relaxation of any element of expanded volume, $\eta\left(z_0,t\right)>0$, is proportional to how long it has been in the expanded state. Numerically, we implement this through the recurrence relation
\begin{equation}\label{eq:relaxtime}
    \tau\left(t,z_0\right)=
    t-\frac{1}{\eta\left(z_0,t\right)}\Big(\eta\left(z_0,t-\Delta t\right)\left[t- \tau\left(z_0,t-\Delta t\right)\right]
    +\lvert\eta\left(z_0,t\right)-\eta\left(z_0,t-\Delta t\right)\rvert \, t\Big), \quad t\geq\Delta t,
\end{equation}
where $\Delta t$ denotes the numerical time increment. That is, the relaxation function $\tau\left(z_0,t\right)$ describes Gaussian relaxation at its core, but is increasingly attenuated the shorter the volume element in question has been in the expanded state. To account for this, the second term in Eq. \eqref{eq:relaxtime} represents the time at which the volume element effectively entered the current expanded state. We achieve this by performing an average weighted by the magnitude of expansion at the previous time step (left-most term in brackets), and the magnitude of the changes in expansion that have occurred since (right-most term in brackets). This protocol naturally resets the relaxation function if the liquid crystal network relaxes back toward its initial configuration, and so permits steady-state oscillations. 
Although it is possible to achieve a similar effect using different protocols, we have chosen our approach under the aspect of simplicity. 
Our relaxation function $\tau\left(z_0,t\right)$ is uniquely determined by temporal evolution of the volume-expansion order parameter profile $\eta\left(z_0,t\right)$, and represents a dynamical coupling that goes beyond the free-energetic nature of the Landau theory.

The above enables us to study the dynamical behavior of the model. However, before we carry out the analysis, we first reduce the parameter space of the model by introducing a dimensionless scaling.

\section{Scaling procedure}\label{sec:scaling}

We scale the theory by introducing the dimensionless volume-expansion order parameter $\tilde{\eta}\equiv\eta/\eta\left(L_0\right)$; we do not scale the population order parameter $f$ [$-$] so that we can explicitly maintain the constraint $0\leq f^2\leq1$. Next, we identify $H_*$ [\SI{}{\joule/\meter^3}] as the relevant energy scale of the problem, and define the scaled field strength $h=\left(H-H_*\right)/H_*$, as well as the scaled bulk modulus $\tilde{B}_f= B_f/H_*$ and coupling constant $\zeta=\xi^2/B_\eta H_*$. Following this, we scale the spatial coordinate $z_0$ to the reference thickness of the film, $\tilde{z}_0=z_0/L_0$, from which the scaled square-gradient coefficients $\tilde{\kappa}_f= \kappa_f/\sqrt{H_*L_0^2}$ and $\tilde{\kappa}_{\tilde{\eta}}=\kappa_\eta/\sqrt{H_*B_{\eta}^2L_0^2/\xi^2}$ follow naturally.

For the scaling of the dynamical model parameters, we opt to set $\tilde{\Gamma}_{\tilde{\eta}}=1$, such that we measure time relative to the evolution of the volume-expansion order parameter $\tilde{\eta}$. It then follows that the time scale of the problem becomes $\tilde{t}=tB_{\eta}^2H_*\Gamma_{\eta}/\xi^2$, and we write the scaled relaxation function $\tilde{\tau}=\tau B_{\eta}^2H_*\Gamma_{\eta}/\xi^2$. Upon inserting these scalings into the dynamical equations, Eq. \eqref{eq: dynamics}, we read off the effective kinetic coefficient of the population order parameter, $\tilde{\Gamma}_f=\Gamma_f\xi^2/\Gamma_{\eta}B_{\eta}^2$, and the effective coefficient for viscoelastic relaxation, $\tilde{\gamma}=\gamma\xi^4/B_{\eta}^4H_*^2\Gamma_{\eta}^2$. For the purpose of this paper, we choose parameter values in the broad regime $\tilde{\Gamma}_f<\tilde{\Gamma}_{\tilde{\eta}}$, which we have previously shown to be qualitatively in line with experiments and Molecular Dynamics computer simulations \cite{kusters2020dynamical}.

This establishes the full theory to be used in the remainder of this paper, which we numerically integrate by means of a straightforward Forward Time Centered Space finite difference scheme. We now discuss the transient dynamics that the model gives rise to.

\section{Transient dynamics}\label{sec:results}

\begin{figure*}[htbp]
    \subfloat[]{\includegraphics[width=8.cm]{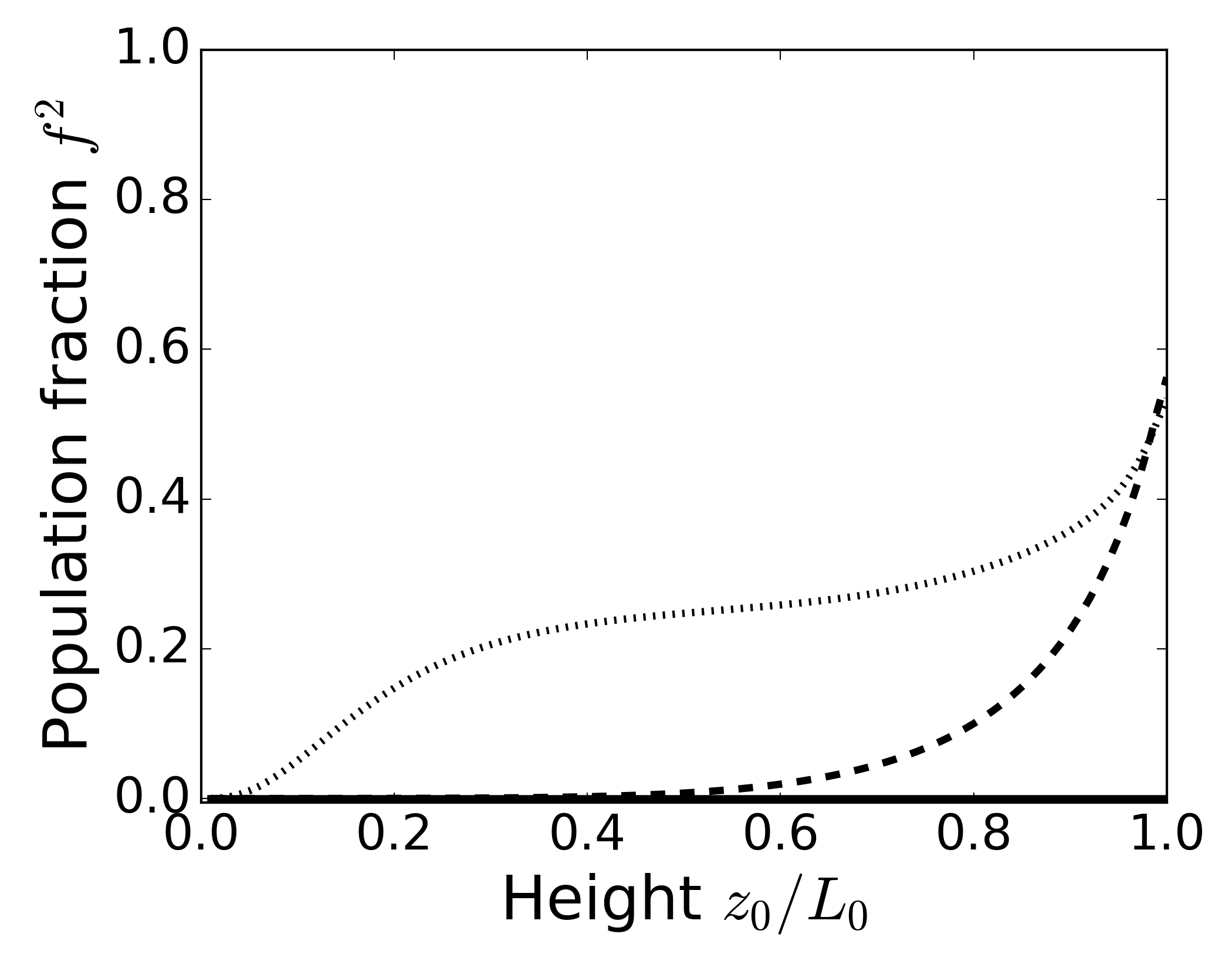}}
    \subfloat[]{\includegraphics[width=8.cm]{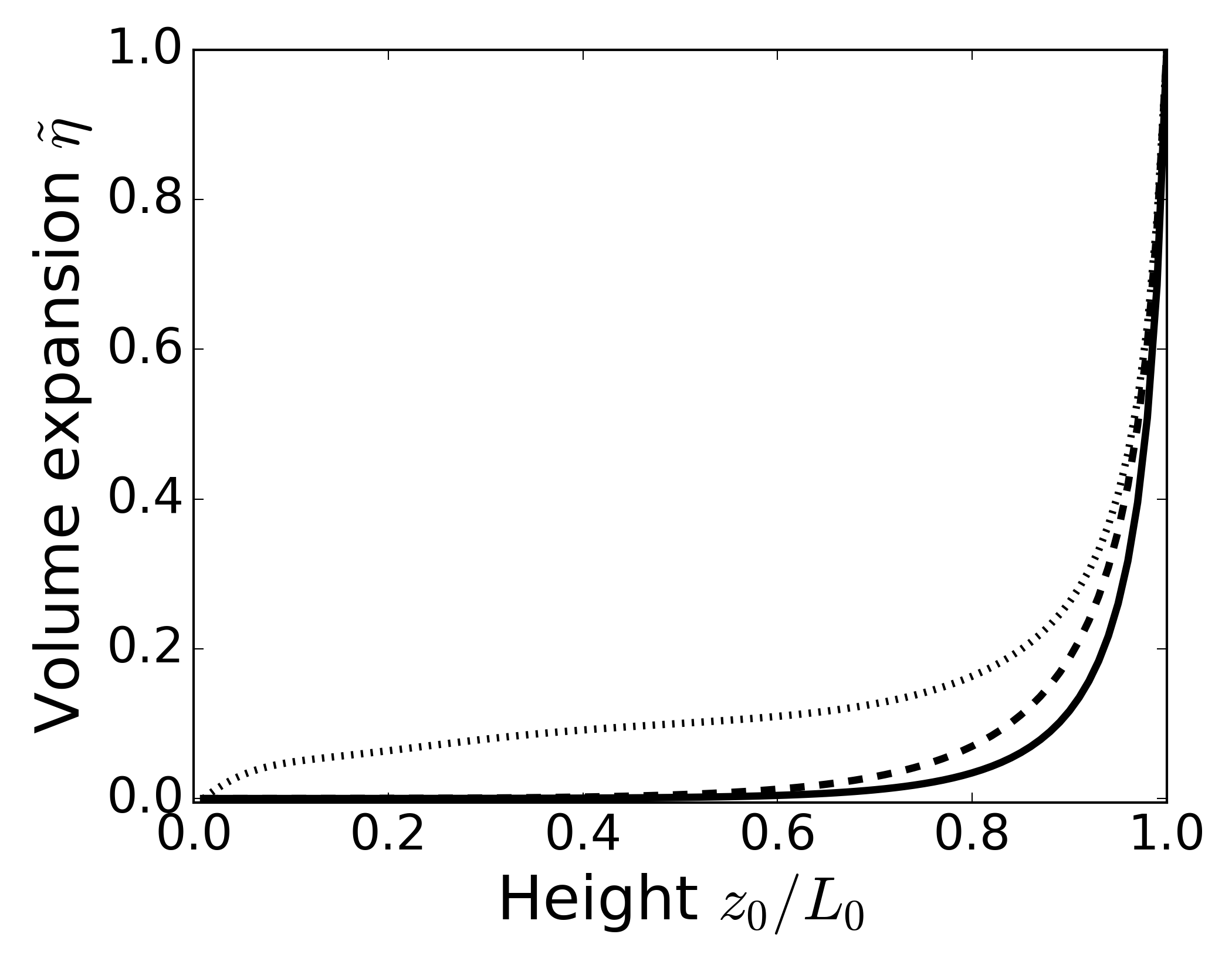}} \\
    \caption{Fraction of mesogens that respond collectively to an applied electric field (a) and the relative volume increase of the liquid crystal network (b), obtained from the scaled theory and as a function of the scaled height coordinate ${z}_0/L_0$. We apply a super-critical alternating electric field, $\left(h+1\right)\cos^22\pi\tilde{\omega}\tilde{t}$, with scaled field strength $h=1.5$ and scaled driving frequency $\tilde{\omega}=0.04$. The solid curves correspond to the initial order parameter profiles ($\tilde{t}=51.4$), whereas the dashed and dotted curves coincide with a saturated response at the top of the film ($\tilde{t}=100.5$) and the steady-state oscillation ($\tilde{t}=140.1$), respectively. See Sec. \ref{sec:scaling} for the scaling procedure. The values used for the scaled model parameters are $\zeta=0.51$, $\tilde{B_f}=3.70$, $\eta\left(L_0\right)=3.375$, $\tilde{\kappa}_f=\tilde{\kappa}_{\tilde{\eta}}=0.071$, $\tilde{\Gamma}_f=0.01$, and $\tilde{\Gamma}_{\tilde{\eta}}=1$.}\label{fig:illustrate_pop_vol}
\end{figure*}

We start our analysis of the transient model dynamics by investigating the temporal evolution of the order parameter profiles in response to an alternating electric field of the form $\left(h+1\right)\cos^22\pi\tilde{\omega}\tilde{t}$, with $\tilde{\omega}=\omega\xi^2/B_{\eta}^2H_*\Gamma_{\eta}$ the scaled driving frequency. Here, the ``$+1$" follows from explicitly writing out the scaling procedure, whereas the square stems from Ref. \cite{Note1}. Fig. \ref{fig:illustrate_pop_vol} (a) shows the population order parameter profile, $f\left(z_0\right)^2$, which provides an indication of the local mesogen response (e.g., collective reorientation), at typical stages during actuation. Our qualitative results turn out to not depend strongly on the driving frequency of the alternating electric field, hence we show only results for a single frequency, $\tilde{\omega}=0.04$. The solid curve (time scale I) shows that the profile, which initially is entirely flat, remains flat for an extended period of time, indicating no appreciable mesogen response exists and in the early stages of actuation. Although it is not apparent from the figure, below we shall see that prior to this time scale the fraction of mesogens coherently reorienting in phase with the electric field already grows exponentially. This becomes \textit{macroscopically} discernible beyond this first time scale, where we find that the growth rate is largest at the free boundary of the film ($z_0/L_0=1$), and, consequently, the response saturates there before a sizable bulk response is realized, as shown by the dashed curve (time scale II). This asymmetry in growth rate stems from the (diffuse) interface with the ambient medium: the amount of available free volume is larger at this boundary compared to the bulk, and thus the mesogens are hampered to a lesser extent in their reorientation. Finally, the response also permeates into the bulk of the thin film and the material enters a steady-state oscillation, as shown by the dotted curve (time scale III). 

The response of the volume-expansion order parameter profile, $\tilde{\eta}\left(z_0\right)$, shown in Fig. \ref{fig:illustrate_pop_vol} (b), follows a similar trend, albeit with an initial profile that is already non-uniform (solid curve, time scale I); this reflects the diffuse interface we model. Moreover, since volume expansion only indirectly responds to the electric field, through coherent mesogen reorientation, we expect that in these early stages of actuation, even on the microscopic scale, no discernible expansion occurs, in contrast to the exponential growth of the population order parameter $f^2$. We illustrate this further below by showing that the response is entirely dominated by thermal noise prior to time scale I. Following this, driven by the mesogen reorientation at the top of the film, as highlighted by the dashed curve in Fig. \ref{fig:illustrate_pop_vol} (a), the film locally expands at the free boundary (dashed curve, time scale II). Finally, the response again permeates the film, giving rise to a steady-state volume expansion (dotted curve, time scale III). Videos of both order parameter profiles are available online as supplemental material \cite{SupplVideo}.

Next, we aim to more closely connect the order parameter profile dynamics with experiments. Since the experimental findings we focus on here primarily concern measurements of surface dynamics \cite{van2019morphing,van2020electroplasticization}, we introduce coarse-grained quantities to represent both order parameters on the macroscopic scale. This results in an integrated measure for the mesogen response, $I_m\left(\tilde{t}\right)=\int_0^{L_0}\diff z_0 \, f^2\left(z_0,\tilde{t}\right)$, which can be probed experimentally by means of Raman scattering or birefringence measurements, as well as for the film thickness, $L\left(\tilde{t}\right)=\int_0^{L_0}\diff z_0 \left[1+\eta\left(z_0,\tilde{t}\right)\right]$. In writing these definitions, and in what follows, we assume that the effect of any local reorientation or deformation instantaneously registers on the macroscopic scale, i.e., such information propagates through the network on time scales much shorter than those under consideration here. This is expected for the thin films used in experiments, and implies that mesogen reorientation and volume expansion can macroscopically result in distinct changes in surface dynamics. Below we discuss the transient dynamics of both sequentially, focussing specifically on the relevant time scales (I, II, and III); for the numerical procedure we use, see Ref. \footnote{Numerically, we measure the time scales as follows. Since we expect the volume-expansion order parameter, $\tilde{\eta}$, to be characterized by thermal noise prior to time scale I, we determine the first time scale by keeping track of the \textit{relative} expansion, $\left(L\left(\tilde{t}\right)-L\left(0\right)\right)/L\left(0\right)$; the time scale then coincides with the last instance of successive time steps differing by more than 5\%. Similarly, time scale II corresponds to a saturated mesogen response at the top of the film and, accordingly, we determine it by identifying the point in time where $f\left(L_0\right)$ first exceeds 95\% of its maximum value. Finally, time scale III demarcates the onset of the steady-state oscillation, which we identify by computing the maximum \textit{relative} expansion the liquid crystal network assumes in the steady state, and finding the first instance where the relative expansion lies between 99\% and 101\% of this value}.

\begin{figure}[htbp]
    {\includegraphics[width=10.cm]{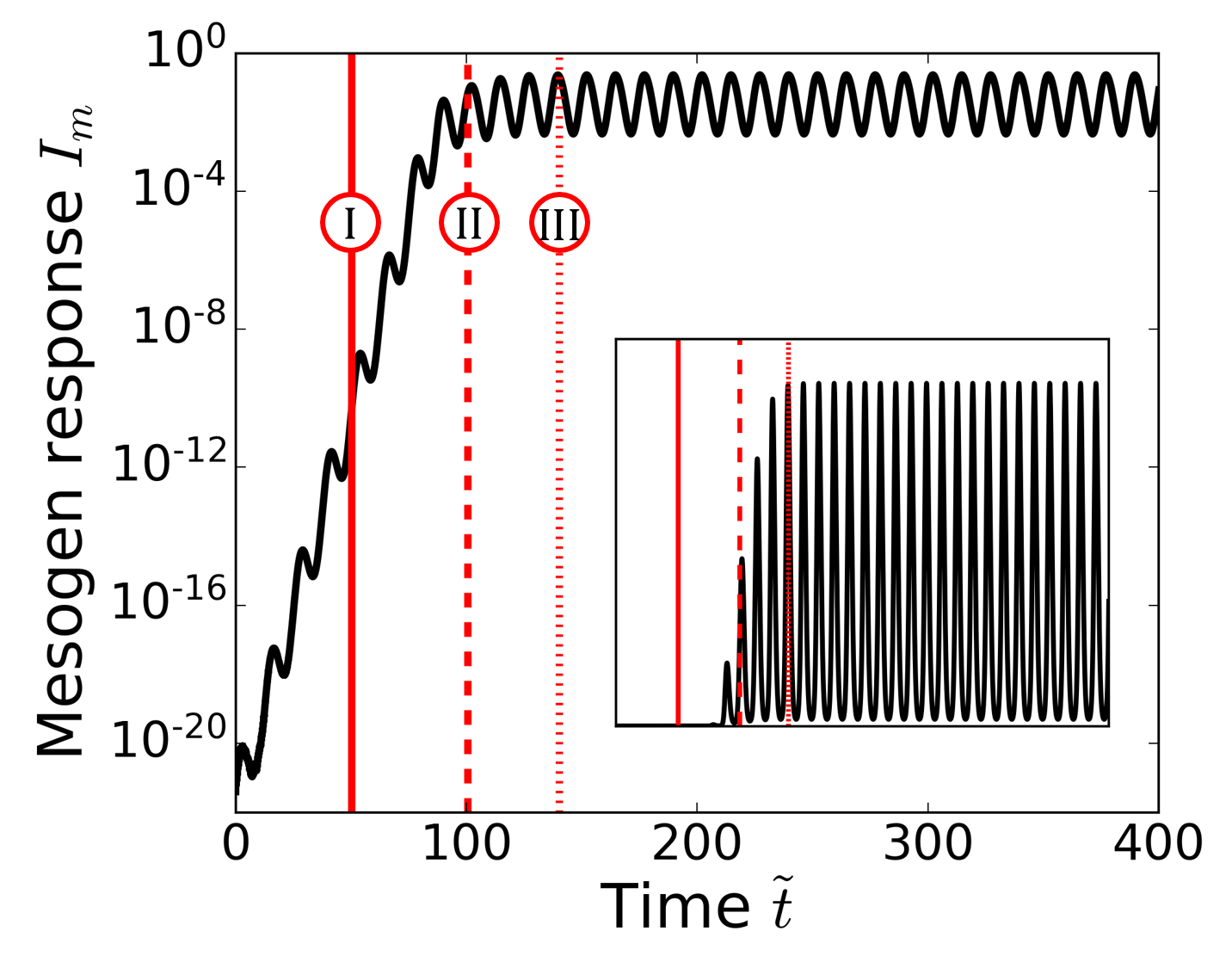}}
    \caption{Overall, integrated mesogen response, $I_m\left(\tilde{t}\right)=\int_0^{L_0}\diff z_0 f^2\left(z_0,\tilde{t}\right)$, as a function of the scaled time $\tilde{t}$, shown on a semi-logarithmic scale. The vertical red lines correspond to the time scales I, II, and III, identified in Fig. \ref{fig:illustrate_pop_vol} and discussed in Sec. \ref{sec:results}, and the inset shows the same figure on a linear scale. See Sec. \ref{sec:scaling} for the scaling procedure and see figure \ref{fig:illustrate_pop_vol} for the parameter values.}\label{fig:illustrate_Im}
\end{figure}

Fig. \ref{fig:illustrate_Im} shows the temporal evolution of the integrated mesogen response, $I_m$, on a logarithmic scale; the inset shows the same curve on a linear scale. From the logarithmic plot we conclude that the overall mesogen response, i.e., the fraction of mesogens coherently reorienting in response to the alternating electric field, initially grows exponentially and subsequently saturates to a steady state. This qualitative behaviour is segmented in the figure by vertical red lines, which coincide with the time scales I, II, and III discussed above, i.e., the order parameter profiles shown in Fig. \ref{fig:illustrate_pop_vol} correspond to the times indicated by the red lines. Prior to the solid red line (time scale I) the mesogen response in the material as a whole grows exponentially, although, as noted above, this is difficult to discern when not looking at the figure on a logarithmic scale. Following this, the growth becomes macroscopically apparent near the top of the film, where the response saturates upon crossing the dashed red line (time scale II); this corresponds to the dashed curve in Fig. \ref{fig:illustrate_pop_vol} (a). Furthermore, the linear-scale inset of Fig. \ref{fig:illustrate_Im} shows that the integrated mesogen response at this time is significantly lower than in the eventual steady state, indicating that the bulk response must also eventually play an important role. Indeed, permeation of the response into the bulk occurs beyond this point, until the dotted red line in Fig. \ref{fig:illustrate_Im} (time scale III) is reached; this effect is most clearly visible in the linear-scale inset. After this, no more qualitative changes in behavior ensue, and the steady-state oscillation is sustained.

\begin{figure}[htbp]
    {\includegraphics[width=10.cm]{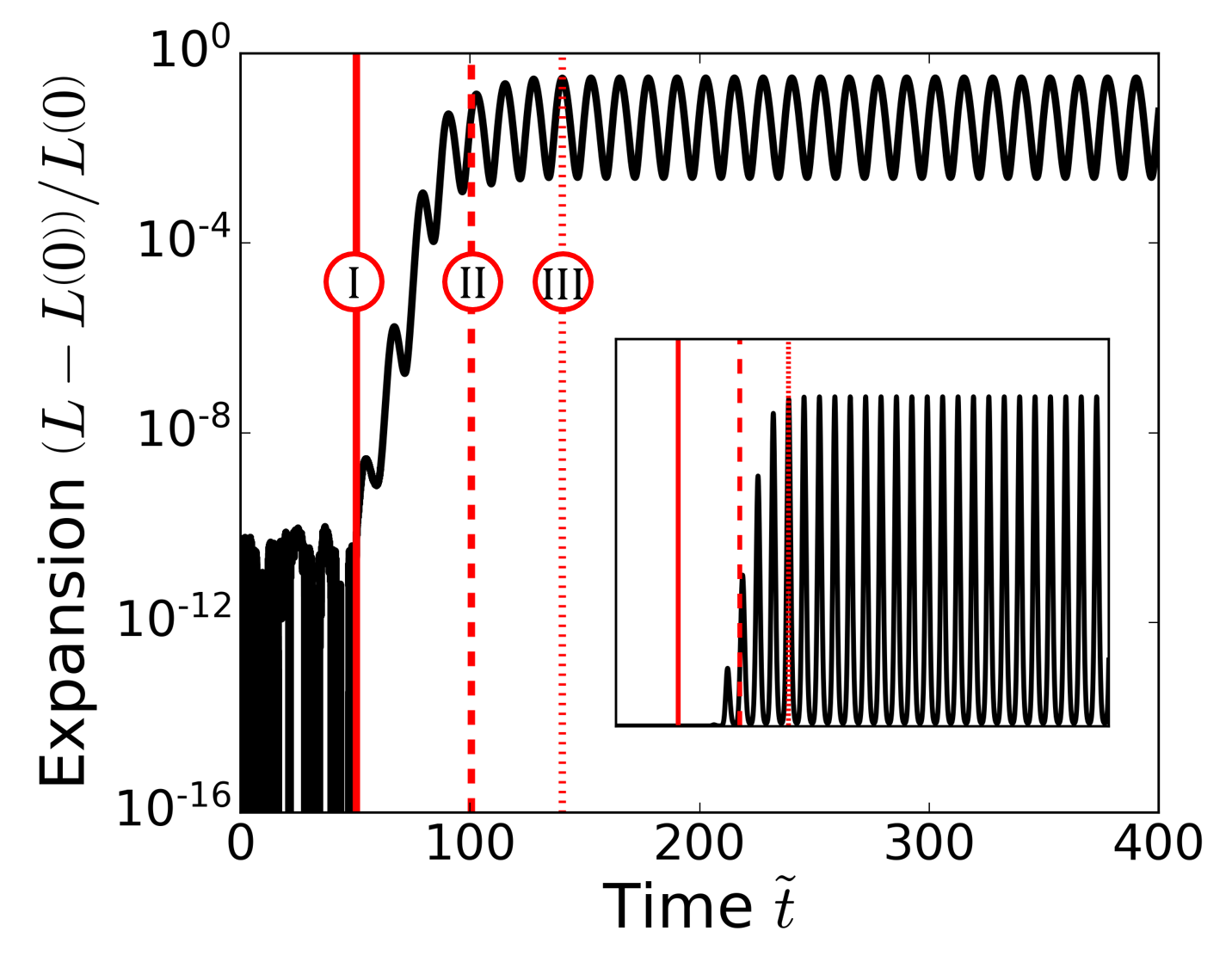}}
    \caption{Expansion of the liquid crystal network film, $\left(L\left(\tilde{t}\right)-L\left(0\right)\right)/L\left(0\right)$, with $L\left(\tilde{t}\right)=\int_0^{L_0}\diff z_0 \left[1+\eta\left(z_0,\tilde{t}\right)\right]$ the film thickness, as a function of the scaled time $\tilde{t}$, shown on a semi-logarithmic scale. The vertical red lines correspond to the time scales I, II, and III, identified in Fig. \ref{fig:illustrate_pop_vol} and discussed in Sec. \ref{sec:results}, and the inset shows the same figure on a linear scale. See Sec. \ref{sec:scaling} for the scaling procedure and see figure \ref{fig:illustrate_pop_vol} for the parameter values.}\label{fig:illustrate_T}
\end{figure}

The concomitant expansion of the film, as shown in Fig. \ref{fig:illustrate_T}, follows the same pattern as described above, with one notable exception: prior to the solid red line (time scale I), the behavior is entirely dominated by noise. As noted above, this is due to the fact that volume expansion is driven by mesogen reorientation, rather than directly through the applied electric field. As a result, the mesogen response must first grow to be strong enough to overcome thermal noise, which occurs as the solid red line is passed. We stress, however, that the expansion associated with this regime is so small in magnitude so as to fall outside the range of validity of our coarse-grained theory, which applies on length scales exceeding the mesh size of the network, i.e., the distance between permanent crosslinks, $l\approx\SI{10}{\nano\meter}$; this length scale shall later return as a system parameter to enable predictions in terms of physical quantities. Below this length scale, the theory provides no meaningful insight, and so we are forced to disregard the corresponding dynamics: instead of directly observing the thermal noise, we recover no discernible response whatsoever. In practice, the time scale at which expansions on the scale of the network mesh size arise lies slightly beyond the solid red line; a similar limit of resolution exists in the experiments due to the limitations of the measurement set-up. After this, the dashed red line (time scale II) again corresponds to a saturated expansion at the top of the film, driven by the local mesogen response, and, finally, the steady state is reached at, and sustained beyond, the dotted red line (time scale III).


Thus, assuming that the features discussed above translate to changes in surface dynamics, we recognize three characteristic time scales, which split up the \textit{transient} dynamics of the material into three distinct regimes. First, prior to time scale I, a collective response of the mesogen builds up; during this regime, no appreciable volume changes are discernible. Secondly, between time scales I and II, we observe a local expansion of the liquid crystal network, focussed at the top of the film; this is driven by local mesogen reorientation, which is impeded to a lesser extent at the free boundary of the film. Finally, between time scales II and III, the response permeates into the bulk of the material, and realizes a steady-state oscillation. This ends the transient stages of dynamics, as beyond this point the steady-state oscillation is merely sustained.

Our findings suggest that permeation of the response into the liquid crystal network film provides an alternative explanation for the three transient regimes Van der Kooij and co-workers report in their experimental work \cite{van2019morphing}. In order to provide a test that distinguishes this interpretation, which implies spatial information translates directly to distinct dynamical behavior on experimentally accessible scales, from the interpretation proposed by Van der Kooij and co-workers, which associates the regimes with the collective response of different mesogen species, we devote the following section to investigating the effect of varying the (initial) film thickness.

\section{Dependence on film thickness}\label{sec:results2}



\begin{figure}[htbp]
    {\includegraphics[width=10.cm]{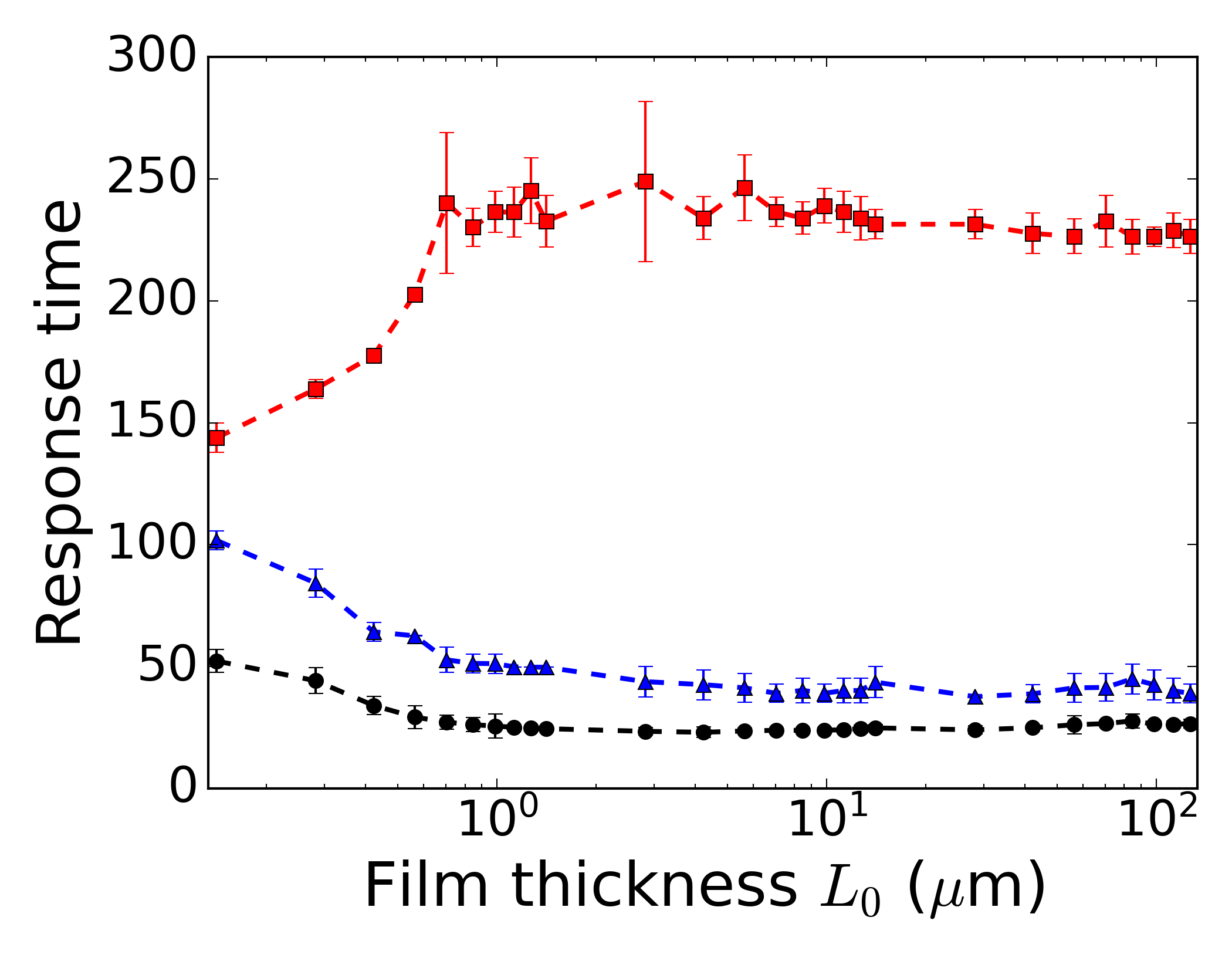}}
    \caption{Time scales corresponding to the different transient regimes shown in Fig. \ref{fig:illustrate_pop_vol} and discussed in Sec. \ref{sec:results}, as a function of the (initial) film thickness, $L_0$, using $l\approx\SI{10}{\nano\meter}$ (see Sec. \ref{sec:results2}). The black curve (circles) corresponds to the time required to overcome thermal noise (lag time), the blue curve (triangles) coincides with a saturated response at the top of the film, and the red curve (squares) denotes the start of the steady-state oscillation. Data points, including error bars, correspond to an average over ten numerical simulations with random thermal noise, and dashed lines are guides to the eye. See Sec. \ref{sec:scaling} for the scaling procedure and see Fig. \ref{fig:illustrate_pop_vol} for the parameter values not explicitly stated here.}\label{fig:regimes}
\end{figure}

In terms of the theory, the initial thickness of the liquid crystal network film enters only through the scaled square-gradient coefficients, $\tilde{\kappa}_f$ and $\tilde{\kappa}_{\tilde{\eta}}$, which depend inversely on it. We shall focus on the former, since the model dynamics is primarily dictated by the population order parameter, $f^2$ \cite{kusters2021permeation}. In particular, the unscaled coefficient, $\kappa_f$ [\SI{}{\sqrt{\joule/\meter}}], represents an energy scale, smeared over a microscopic length scale of the system. To estimate this parameter, we point out that the scaled field strength, $H_*$ [\SI{}{\joule/\meter^3}], is the relevant energy scale for mesogen reorientation, whereas we have both a length scale associated with the mesh size of the network, $l$ [\SI{}{\meter}], and a length scale associated with the macroscopic thickness of the film $L_0$ [\SI{}{\meter}]. Since only the former is relevant on the microscopic scale, we approximate $\kappa_f\approx \sqrt{H_*}l$, where we shall use $l\approx\SI{10}{\nano\meter}$. The scaled coefficient then makes the dependence on the (initial) film thickness explicit through the relation $\tilde{\kappa}_f=\kappa_f/\sqrt{H_*}L_0$; for the sake of simplicity, we set $\tilde{\kappa}_{\tilde{\eta}}=\tilde{\kappa}_f$, as both scale identically with the film thickness. We thus vary $\tilde{\kappa}_f=l/L_0$ as a proxy for the film thickness, indicating the ratio between the microscopic and macroscopic length scales of the system is the relevant control parameter. Figure \ref{fig:regimes} shows the resulting transient time scales, averaged over ten numerical simulations with random thermal noise, $\delta$-correlated in both time and space.



In the figure, the black curve corresponds to the time required to overcome the thermal noise (regime I), the blue curve coincides with a saturated response at the top of the film (regime II), and the red curve denotes the time at which the material enters its steady-state oscillation (regime III). Reading the figure from left to right, it becomes apparent that a thin-film-to-bulk transition occurs, approximately, at $L_0=\SI{1}{\micro\meter}$ if we presume $l\approx\SI{10}{\nano\meter}$. Upon traversing this transition, the times required to overcome thermal noise and achieve a saturated response at the top of the film decrease sharply, whereas the time required to achieve a steady-state oscillation increases. We rationalize the former two by noting that decreasing the square-gradient coefficients weakens the local interactions within the model, i.e., if we make the film thicker, the range of mesogen-mesogen interactions shrinks relative to the thickness of the film. As a result, the inhibiting effect of the substrate onto which the film is clamped effectively affects an increasingly small proportion of the film. Consequently, a greater proportion of mesogens can reorient freely in response to an applied electric field, and the response accelerates. We attribute the fact that the steady-state oscillation nevertheless takes longer to establish in this case to the onset of domain formation, which we shall demonstrate in the following section; until then, we treat this as a working hypothesis. Finally, we note that although there exists an additional transition for exceedingly large film thicknesses (on the centimeter scale, results not shown), these are not relevant length scales for the smart coatings we consider here, and so fall outside the scope of this paper.

\begin{figure}[htbp]
    {\includegraphics[width=10.cm]{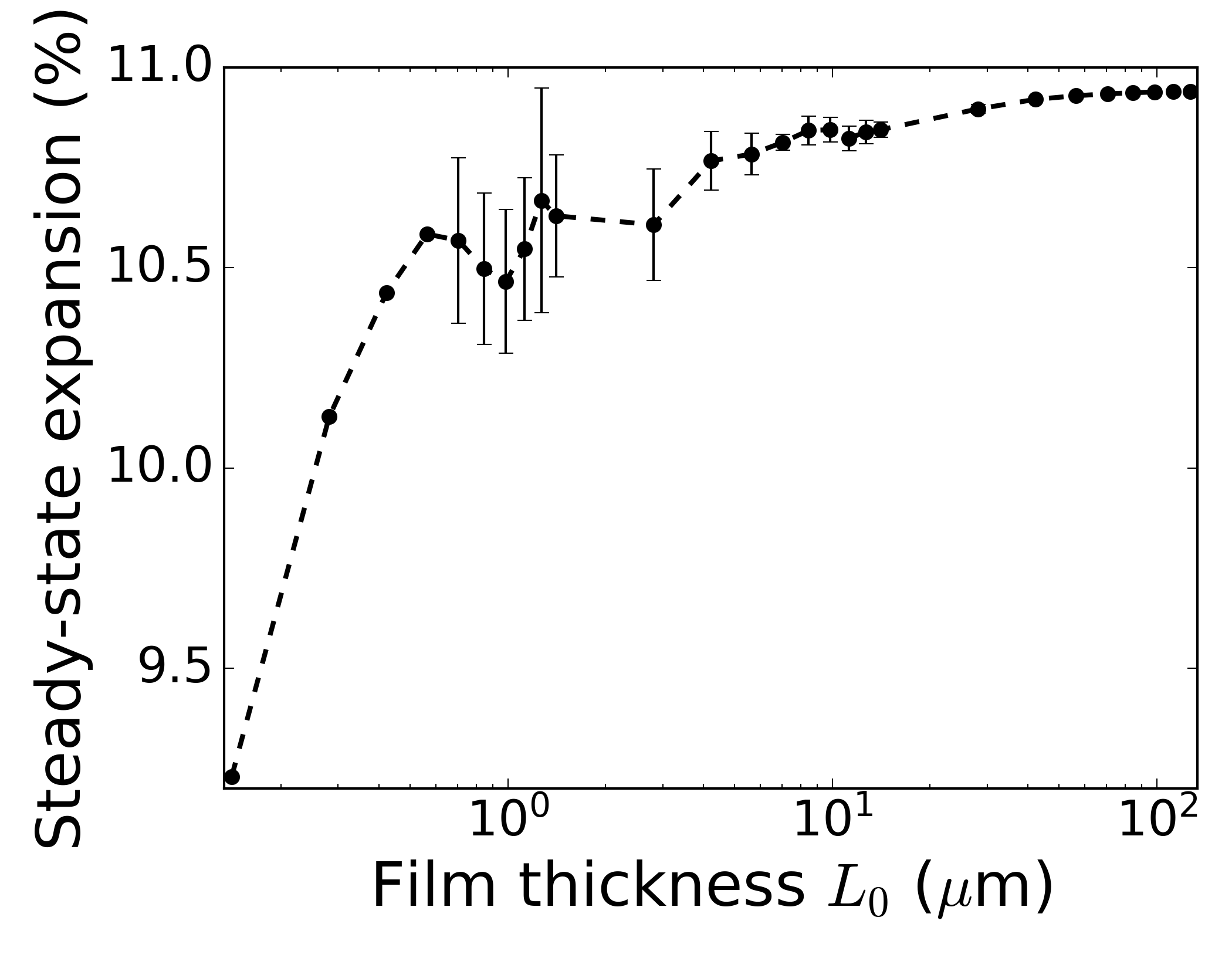}}
    \caption{Steady-state expansion of the liquid crystal network film, as a function of the (initial) film thickness, $L_0$. Data points, including error bars, correspond to an average over ten numerical simulations with random thermal noise, and dashed lines are guides to the eye. See Sec. \ref{sec:scaling} for the scaling procedure, and see Figs. \ref{fig:illustrate_pop_vol} and \ref{fig:regimes} for the parameter values not explicitly stated here.}\label{fig:regimes volume}
\end{figure}

The experimental implications of varying the film thickness are broader, however, than just altering the transient time scales of the liquid crystal network film. Fig. \ref{fig:regimes volume} illustrates that the steady-state volume expansion of the film is also affected. In particular, we find that in the thin-film limit (left), increasing the film thickness increases the magnitude of expansion, which approaches a constant value in the bulk limit (right). Furthermore, operation of the material becomes less predictable near the thin-film-to-bulk transition, around $L_0=\SI{1}{\micro\meter}$, as evidenced by the large error bars. This implies that a large spread of steady-state expansion can be expected when operating the liquid crystal network near the thin-film-to-bulk transition in experiments. Thus, there is a trade-off to be made, where quick steady-state actuation can be achieved in thin films at the expense of the overall deformation magnitude, whereas thick films achieve a greater deformation at the expense of more drawn-out transient dynamics. The intermediate region should be avoided due to its unreliability.

This concludes our discussion of the effect the initial film thickness has on the transient dynamics. We point out that a similar dependence on initial film thickness is not expected if one assumes the different transient regimes result from the collective reorientation of different mesogen species, as Van der Kooij and co-workers propose \cite{van2019morphing}. Accordingly, experiments on films with varying thicknesses can be used to further probe the underlying mechanism. Finally, as announced above, we demonstrate that the thin-film-to-bulk transition is indeed characterized by the onset of domain formation, which contextualizes our results.

\section{Domain formation}\label{sec:results3}

\begin{figure*}[htbp]
    \subfloat[]{\includegraphics[width=8.cm]{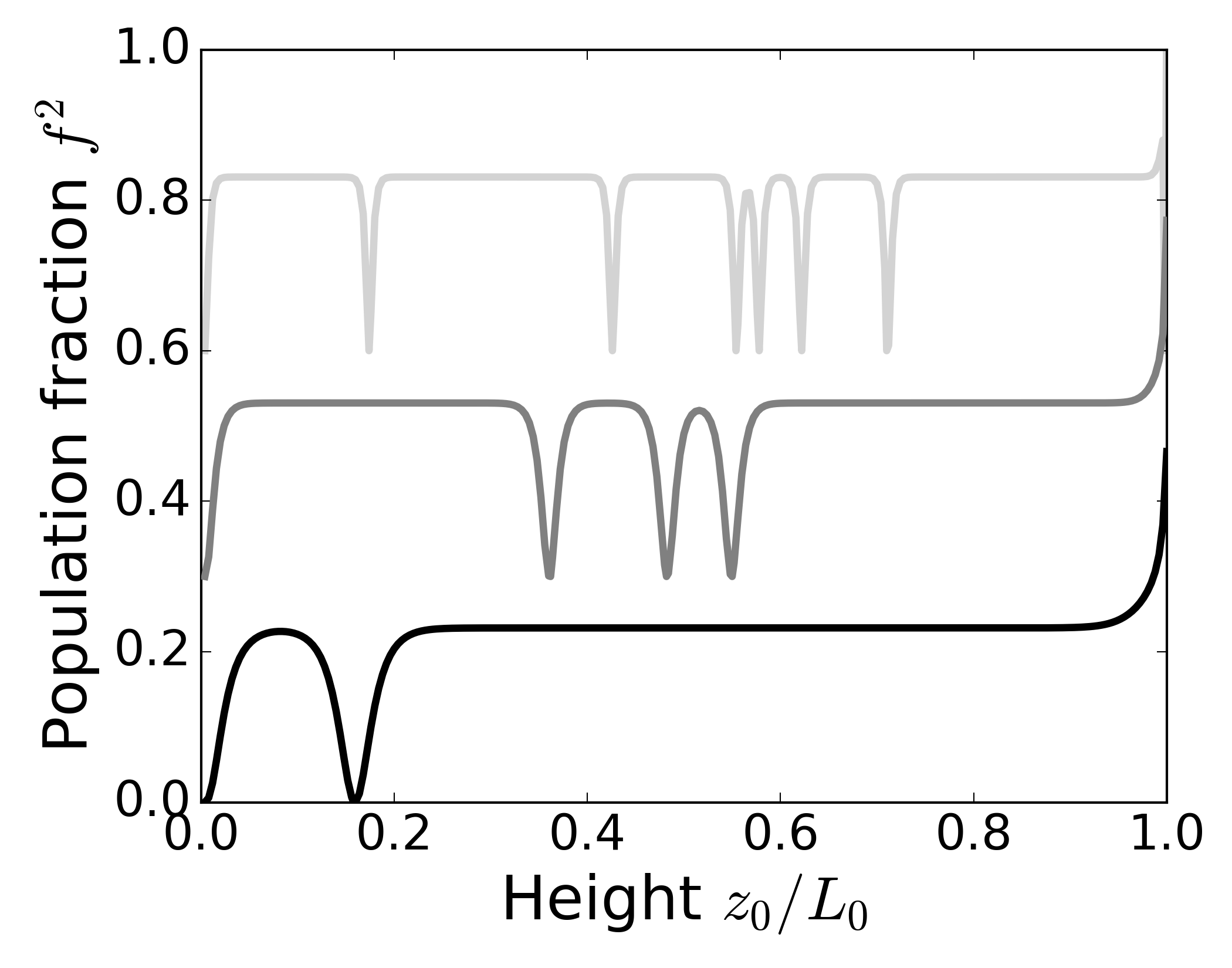}}
    \subfloat[]{\includegraphics[width=8.cm]{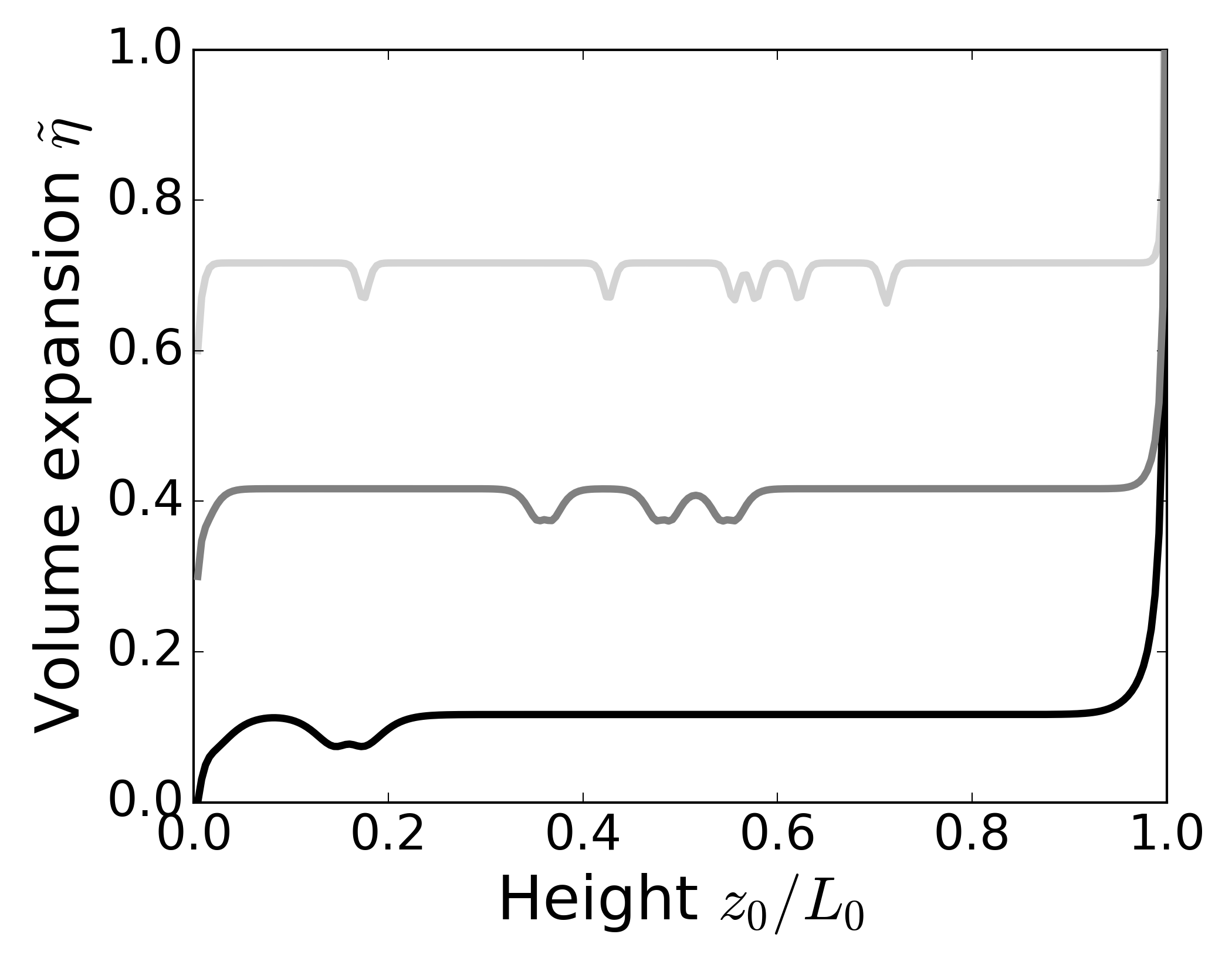}} \\
    \caption{Volume-expansion order parameter, $\tilde{\eta}$, as a function of the scaled height coordinate ${z}_0/L_0$. The different curves are offset by multiples of 0.3 for visual clarity, and correspond to the different (initial) film thicknesses $L_0=\SI{1}{\micro\meter}$ (black), $L_0=\SI{2}{\micro\meter}$ (gray), and $L_0=\SI{5}{\micro\meter}$ (light gray). See Sec. \ref{sec:scaling} for the scaling procedure, and see Figs. \ref{fig:illustrate_pop_vol} and \ref{fig:regimes} for the parameter values not explicitly stated here.}\label{fig:domains profiles}
\end{figure*}

Fig. \ref{fig:domains profiles} shows the steady-state order parameter profiles for various values of the (initial) film thickness, $L_0$, offset by multiples of 0.3 for visual clarity. From this it is apparent that around $L_0=\SI{1}{\micro\meter}$ (black curve) the first domains form, which are separated by a domain wall, i.e., a small region in which, locally, the degree of mesogen reorientation and volume expansion are significantly smaller than around it. The small bump that is visible at the center of the domain wall in Fig. \ref{fig:domains profiles} (b) stems from the fact that here the gradient in mesogen reorientation is largest. Such energetically unfavorable gradients can be smeared out by locally expanding the material, i.e., regions in which the fraction field-aligned mesogens varies significantly generally require more space than regions where this fraction varies weakly. Thus, this bump is a direct consequence of the geometric significance of the volume-expansion order parameter $\tilde{\eta}$.

Upon increasing the (initial) film thickness, $L_0$, the number of domains increases accordingly, as shown by the gray ($L_0=\SI{2}{\micro\meter}$) and light gray curves ($L_0=\SI{5}{\micro\meter}$). Simultaneously, the domain walls separating the different domains become narrower, as the effective range of local interactions shrinks relative to the thickness of the film. Note that the positions of these domain walls are arbitrary, as the total free energy is invariant under their translation. This suggests repeated numerical simulations should yield domain walls at different positions, which indeed they do. In addition, it implies domain wall motion as a function of time, although this process becomes negligibly slow for the small ratio of noise to the characteristic energy scale of the problem, i.e., the critical field strength, $H_*$, we use, combined with the limited spatial resolution of our numerical implementation.

\begin{figure}[htbp]
    {\includegraphics[width=10.cm]{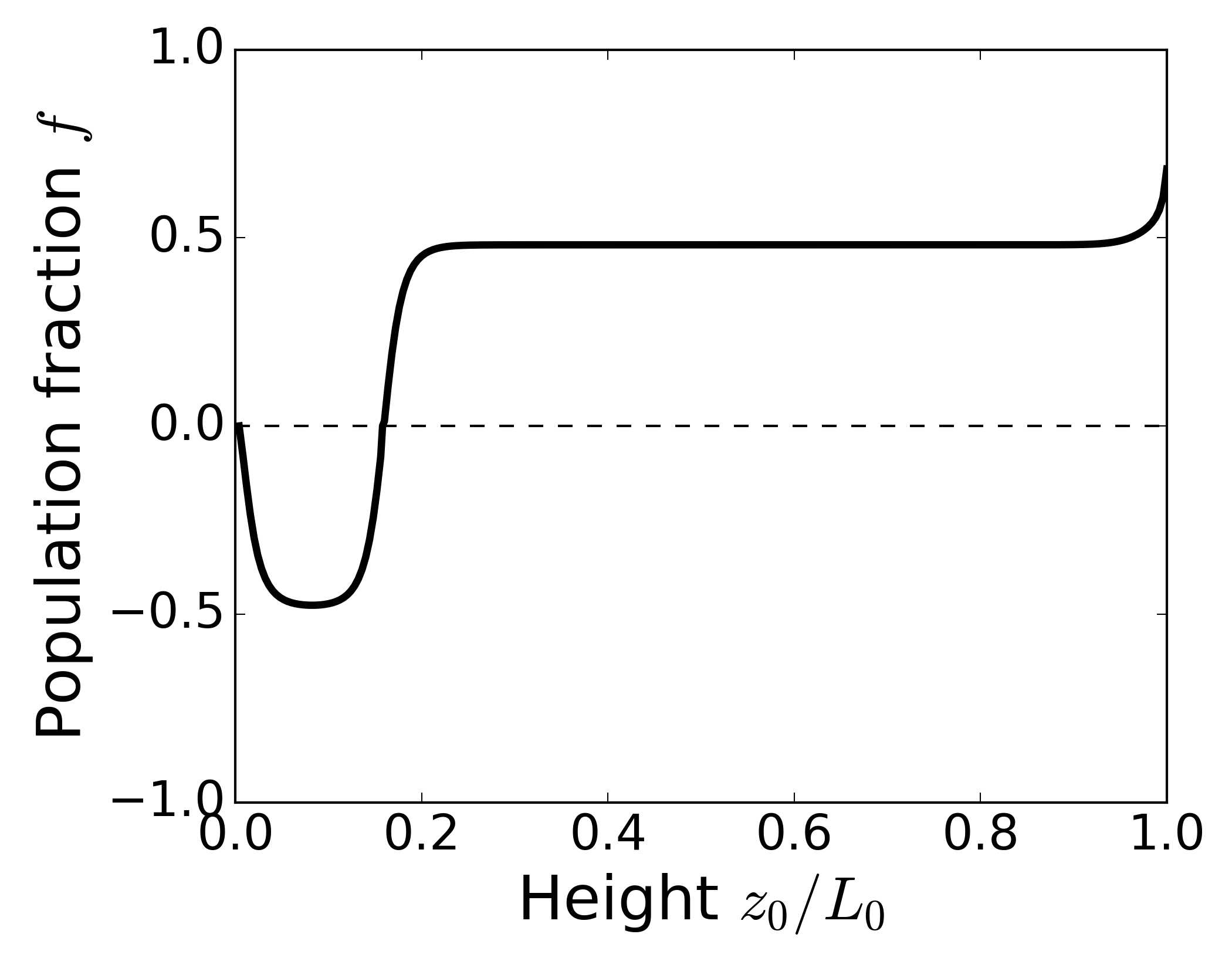}}
    \caption{Population order parameter profile, $f$, as a function of the scaled height coordinate ${z}_0/L_0$. We can interpret the sign of $f$ as indicative of different modes of mesogen reorientation. See Sec. \ref{sec:scaling} for the scaling procedure, and see Figs. \ref{fig:illustrate_pop_vol} and \ref{fig:regimes} for the parameter values not explicitly stated here.}\label{fig:domains profilesf}
\end{figure}

Mathematically, we can interpret the different domains by more closely inspecting the population order parameter. Although only $f^2$ couples to volume expansion, and so is the physical quantity of interest, representing the fraction of coherently reorienting mesogens, in our dynamical equations (Eq. \eqref{eq: dynamics}) we explicitly compute the value of $f$ to recover this quantity. This non-squared order parameter profile contains information that is lost upon squaring, namely its sign, which Fig. \ref{fig:domains profilesf} shows is opposite in neighboring domains. Thus, one way of interpreting different regimes is by associating different signs of $f$ with different, but equi-free-energetic, modes of mesogen reorientation. Since, within the model assumptions, parallel and antiparallel alignment of the dipolar mesogens with the electric field carry the same free energy, a possible candidate for this would be coherent mesogen reorientation in phase and in antiphase with the field. That is, in neighboring domains the mesogens oscillate coherently in opposite directions, with the oscillation being smothered in the domain wall between them, due to their mutual (excluded-volume) interactions.

The fact that domain formation occurs at all suggests that permeation of the response is sufficiently slow so as not to permeate the material before the bulk response ensues, i.e., the bulk response is not absorbed by a permeating front. As a result, it takes longer for the steady-state oscillation to be established (see Fig. \ref{fig:regimes}), since the bulk response must grow virtually unaided by permeation. 



In addition, we likewise attribute the variable steady-state expansion of the liquid crystal network near the thin-film-to-bulk transition (see Fig. \ref{fig:regimes volume}) to the onset of domain formation, as the number of formed domains can vary appreciably depending on the realization of thermal noise. The reason that the steady-state expansion nevertheless reaches a stable plateau in the bulk limit, which is characterized by an even larger number of domains, we trace back to the decreasing width of the corresponding domain walls. That is, since the walls separating different domains become negligibly small compared to the thickness of the film, the effect of domain formation on the achieved steady-state expansion becomes negligible as well.

\begin{figure}[htbp]
    {\includegraphics[width=10.cm]{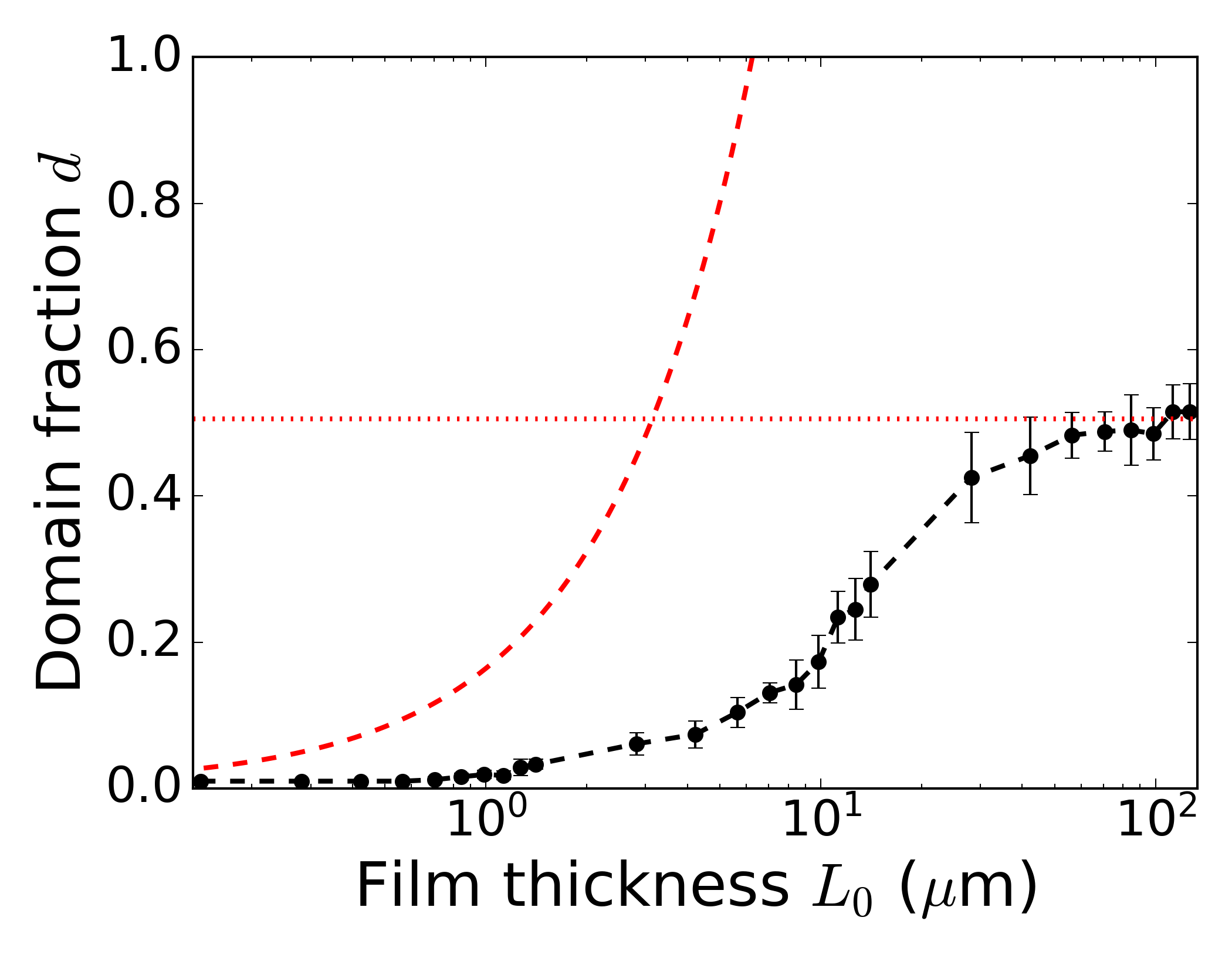}}
    \caption{Number of domains per lattice site used in our numerical scheme, $d$, as a function of the (initial) film thickness, $L_0$. Black data points, including error bars, correspond to an average over ten numerical simulations with random thermal noise, and the dashed black lines are guides to the eye. The dashed red curve indicates the limit of stability following a simple linear stability analysis (see Eq. \eqref{eq:domain}), and the dotted red line denotes the asymptotic limit $d=1/2$. See Sec. \ref{sec:scaling} for the scaling procedure, and see Figs. \ref{fig:illustrate_pop_vol} and \ref{fig:regimes} for the parameter values not explicitly stated here.}\label{fig:domains}
\end{figure}

Finally, we quantify our claims regarding domain formation by computing the average number of domains as a function of the square-gradient coefficient, $\tilde{\kappa}_f$, functioning as proxy for the initial thickness of the film. Fig. \ref{fig:domains} shows this number of domains, normalized by the number of lattice sites used in our numerical scheme, i.e., it shows the average number of domains per lattice site. The black data points show that the onset of domain formation indeed occurs around $L_0=\SI{1}{\micro\meter}$, beyond which it steadily grows until it saturates around $L_0=\SI{100}{\micro\meter}$. This is also roughly the point at which both the time scales corresponding to the different transient regimes, as well as the magnitude of steady-state volume expansion, plateau.

We can straightforwardly understand the asymptotic value the domain fraction, $d$, assumes by considering that in the bulk limit (right) local interactions are negligible. Consequently, the problem of determining the number of expected domains is equivalent to determining the number of clusters in an array of ideal, non-interacting spins in one dimension. For a large array size $N\gg1$, edge effects become negligible and each subsequent spin has an equal probability of introducing an additional domain or not. This suggests the domain fraction must approach $d=1/2$ in the bulk limit, as indicated by the dotted red line.

Furthermore, although the approach toward this asymptotic value is significantly more complicated in terms of both energetic and entropic considerations, we provide a rough estimate on the qualitative behavior. This we do by investigating the linear stability of infinitesimal perturbations of the form $\sin\frac{2n-1}{2}\pi z_0$, where $n=1,2,3,\dots$ represents the number of domains; these perturbations obey the boundary conditions by construction. For the purpose of this calculation, we consider only the population order parameter, $f^2$, as it dominates the (linear) model dynamics \cite{kusters2021permeation}. We solve the resulting dispersion relation to identify which modes become unstable and which damp out, yielding
\begin{equation}\label{eq:domain}
    n=\frac{1}{\pi\tilde{\kappa}_f}\sqrt{\frac{h-1}{2}}+\frac{1}{2}
\end{equation}
as the limit of stability. Here, we made our results independent of the driving frequency, $\tilde{\omega}$, by assuming the alternating electric field oscillates sufficiently quickly for the liquid crystal network to effectively `feel' only the average field, $\left(h+1\right)\cos^2{2\pi\tilde{\omega} \tilde{t}}\rightarrow \left(h+1\right)/2$. Although this estimate, shown by the dashed red line in Fig. \ref{fig:domains}, clearly overestimates the expected domain fraction, neglecting the explicit growth rates, non-harmonic modes, lower-order modes, and the number of configurations possible with each domain fraction, it does provide a qualitative basis for the onset of domain formation. In fact, the curve can be fitted quite well to the numerical data by means of a simple scaling factor, suggesting Eq. \eqref{eq:domain} fundamentally captures the same underlying physics as the full model. 

The above suggests that the scaled model parameters that appear in Eq. \eqref{eq:domain} can be used to temper the formation of domains. In Sec. \ref{sec:results}, we already argued that the scaled square-gradient coefficient obeys the approximate relation $\tilde{\kappa}_f\approx l/L_0$, with $l$ the mesh size of the network and $L_0$ the (initial) film thickness. Similarly, we characterize the scaled field strength, $h=\left(H-H_*\right)/H_*$, by pointing out that the energy of a mesogen with a permanent dipole moment, $\underline{p}$, subject to the electric field, $\underline{E}$, scales crucially with the mesogen orientation, such that we recover $H\propto S_0\lvert\underline{p}\rvert^2\lvert\underline{E}\rvert^2$ for the orienting field, with $S_0$ the orientational order of the mesogens crosslinked into the polymer matrix \cite{kusters2020dynamical,Note1} 
. The excluded-volume interactions opposing reorientation likewise depend strongly on the orientations of the mesogens, as well as their spatial dimensions, according to $H_*\propto S_0^2a^2r$, where we have assumed identical rod-like mesogens with length $a$ and radius $r$ \cite{kusters2020dynamical,onsager1949effects}. Subsequently recognizing that the mesh size of the network, $l$, and the dipole moment of pendant mesogens, $\lvert\underline{p}\rvert$, must, all else being equal, also scale with the mesogen length, and inserting the result into Eq. \eqref{eq:domain}, recover
\begin{equation}\label{eq:domain2}
    n=\frac{c_1L_0}{a+c_2}\sqrt{\frac{1}{S_0r}-c_3}+\frac{1}{2},
\end{equation}
with $c_1, c_2, c_3$ phenomenological fitting parameters. We thus conclude that the formation of domains can be suppressed by decreasing the initial thickness of the film, or by increasing either the linear dimensions of the mesogens, or the orientational order with which they are crosslinked into the network. We stress, however, that these modifications generally come at the cost of a decreased deformation magnitude, meaning a trade-off is again unavoidable.

\section{Conclusion and discussion}\label{sec:conclusion and discussion}
In summary, we investigated the transient dynamics of electrically-deforming liquid crystal network films, inspired by the experiments of Van der Kooij and co-workers \cite{van2019morphing}, who report three distinct transient regimes. Although they argue these regimes originate from the sequential response of dipolar and crosslinked mesogens, followed by a progressive weakening of the polymer matrix, no concrete modeling is provided. We propose an alternative explanation of their results in terms of permeation, where the response must initially overcome thermal noise, then saturates at the top of the film, and finally permeates into the bulk. We point out that our choice of reflecting boundary condition for the population order parameter, $\partial_{z_0}f\left(L_0\right)=0$, is crucial to recovering these results, as a Dirichlet boundary condition such as $f\left(L_0\right)=1$, like we used in a previous publication \cite{kusters2021permeation}, would initialize the top with an already saturated response, leading to immediate permeation. Although this was sufficient for our aims in this prior publication, namely the study of permeation, such an initialization is evidently ad hoc. The boundary condition we use in this paper does not suffer from this, as it presupposes no initial response in the absence of an electric field. Instead, this boundary condition prevents order parameter flow through the top of the film, beyond which the mesogen response has no physical meaning. Hence it is the more logical choice for the study of \textit{transient} dynamics.

We show that the time scales characterizing the transient dynamics depend strongly on the initial thickness of the film, giving rise to a thin-film-to-bulk transition that can, if observed, distinguish our interpretation from the one proposed by Van der Kooij and co-workers. The initial thickness of the film likewise has a large effect on the magnitude of steady-state expansion, which increases with film thickness, although this comes at the cost of establishing the steady-state oscillation more slowly. Notably, the expansion magnitude exhibits large error bars at the thin-film-to-bulk transition, making this regime unsuitable for stable actuation.

Finally, we contextualize our findings by pointing out that the thin-film-to-bulk transition coincides with the onset of domain formation, which rationalizes both the slower establishment of a steady-state oscillation, and the less predictable actuation at the transition. Furthermore, using simple linear-stability arguments, we infer that domain formation can be suppressed by, in addition to decreasing the initial thickness of the film, increasing the linear dimensions of the mesogens, and the orientational order with which they are crosslinked into the network.
 
To put our results into perspective, we reiterate that a direct comparison with the experiments of Van der Kooij and co-workers \cite{van2019morphing} requires the assumption that a local response of the liquid crystal network, be it mesogen reorientation or expansion, directly translates to distinct changes in surface dynamics, which are measured experimentally. In addition, we remind the reader that, in the experiments, the time scale of actuation (hundreds of kilohertz) is many orders of magnitude shorter than the time required for the material to enter the steady state (tens of seconds). Although we find these time scales to only differ by approximately an order of magnitude for the driving frequency used in this paper, this mismatch of time scales quickly increases as we further increase the driving frequency; this is because our results do not depend strongly on the driving frequency. Such a discrepancy need not necessarily be grounds for worry in the context of a generic, Landau-type theory, as a host of complex mechanisms is effectively abstracted into a relatively simple, two-order-parameter theory (see also Ref. \cite{kusters2020dynamical}). With this paper, we have nevertheless made explicit the advantages of such a framework in providing a concrete basis for arguments, as well as testable predictions. As such, our work contributes to the realization of finer control over, and greater understanding of, the time scales on which liquid crystal network materials can be activated.

\section*{Author Contributions}
Guido L. A. Kusters: conceptualization, formal analysis, investigation, methodology, software, validation, visualization, writing - original draft. Paul van der Schoot: conceptualization, funding acquisition, resources, supervision, writing - review \& editing. Cornelis Storm: conceptualization, funding acquisition, resources, supervision, writing - review \& editing.

\section*{Conflicts of interest}
There are no conflicts to declare.

\section*{Acknowledgements}
This research received funding from the Dutch Research Council (NWO) in the context of the Soft Advanced Materials (SAM) consortium in the framework of the ENW PPP Fund for the top sectors and from the Ministry of Economic Affairs in the framework of the `PPS-Toeslagregeling'.


\bibliography{Bibliography}

\end{document}